\documentclass{mn2e}
\usepackage{graphicx,times}

\renewcommand{\textfraction}{0.4}

\def \kms {${\rm{km}\,\rm{s}^{-1}}$}

\title[The stellar initial mass function in red-sequence galaxies]
{The stellar initial mass function in red-sequence galaxies:\\1$\mu$m spectroscopy of Coma Cluster galaxies with 
Subaru/FMOS\thanks{Based on data collected at Subaru Telescope, which is operated by the National Observatory of Japan.}}
\author[Russell J. Smith et al.]
{Russell J. Smith$^{1}$\thanks{Email: russell.smith@durham.ac.uk}, John R. Lucey$^{1}$ \& David Carter$^{2}$
~\\
$^1$Department of Physics, University of Durham, Science Laboratories, South Road, Durham DH1 3LE\\
$^2$Astrophysics Research Institute, Liverpool John Moores University, Twelve Quays House, Egerton Wharf, Birkenhead CH41 1LD
}

\date{Accepted 2012 August 10.  Received 2012 August 7; in original form 2012 June 17}

\pagerange{\pageref{firstpage}$-$\pageref{lastpage}} \pubyear{2012}

\begin{document}

\label{firstpage}

\maketitle

\begin{abstract}
To investigate possible variations in the stellar initial mass function (IMF) in red-sequence galaxies, we have obtained 
infrared spectroscopy with Subaru/FMOS for a sample of 92 red-sequence galaxies in the Coma cluster. 
Velocity dispersions, ages and element abundances for these galaxies have been previously determined from optical data.
The full range of velocity dispersions covered by the sample is $\sigma$\,=\,50--300\,\kms. 
By stacking the FMOS spectra in the rest frame, removing sky-subtraction residuals and other artefacts fixed in the observed frame,
we derive composite spectra in the 9600--10500\,\AA\ range for galaxies grouped according to their velocity dispersion or Mg/Fe ratio.
We measure the Wing--Ford band of FeH and a new index centred on a Ca{\,\sc i} line at 10345\,\AA; 
these features are strong in cool dwarf stars, and hence reflect the form of the IMF at low mass ( $<$\,0.5\,$M_\odot$).
The Ca{\,\sc i} line, unlike the Wing--Ford band and other ``classical'' IMF indicators (Na{\,\sc i} doublet, Ca{\,\sc ii} triplet),
is unaffected by the abundance of sodium. 
We compare the measured indices against predictions from spectral synthesis models matched to the element abundances estimated 
from the optical data.
Binning galaxies by velocity dispersion, we find that both the Wing--Ford and Ca{\,\sc i} index measurements are best 
reproduced by models with the Salpeter IMF. There is no clear evidence for an increase in dwarf-star content with velocity dispersion
over the range probed by our sample (which includes few galaxies at the highest velocity dispersions, $\sigma>250$\,\kms).
Binning the observed galaxies instead by Mg/Fe ratio, the behaviour of both indices implies a trend of IMF from Chabrier-like, at abundance ratios close to 
solar, to Salpeter or heavier for highly $\alpha$-enhanced populations. At face value, this suggests that the IMF depends on the 
mode of star formation, with intense rapid star-bursts generating a larger population of low-mass stars.
\end{abstract}

\begin{keywords}
galaxies: elliptical and lenticular, cD --- galaxies: stellar content --- stars: luminosity function, mass function
\end{keywords}

\renewcommand{\textfraction}{0.8}

\section{Introduction}\label{sec:intro}

The stellar initial mass function (IMF) is a crucial ingredient in interpreting extragalactic observations: 
without constraints, or assumptions, for the IMF, observed luminosities cannot be converted into estimates of assembled stellar mass. 
For old galaxies with a Salpeter (1955) IMF, the great majority of output luminosity is contributed by giant-branch and 
main-sequence stars with mass $\sim$1$M_\odot$. 
However, some 80 per cent of the stellar mass is locked up in cool dwarfs with mass $<$0.5$M_\odot$, which collectively provide less than ten per cent of the 
bolometric luminosity. The total stellar mass-to-light ratio is hence very sensitive to the form of the IMF at low masses. 

In the Milky Way, resolved star counts indicate that the IMF follows a Salpeter-like power law ${\rm d}N(M)$\,$\propto$\,$M^{-2.35}{\rm d}M$ for $M$\,$\ga$\,$M_\odot$, 
but becomes shallower at lower masses. This form can be well represented either by a broken power law (e.g Kroupa, Tout \& Gilmore 1993), or by a log-normal distribution 
(Chabrier 2003). Extensive searches for variation in the IMF in different environments within the Milky Way have not found conclusive evidence for
non-universality (e.g. see the review by Bastian, Covey \& Meyer 2010).
For extragalactic systems, low-mass stars are not directly observable, and the application of a Chabrier-like IMF to 
distant galaxies must be regarded with caution until empirically tested. 
Limits on the contribution of dwarf stars to the mass budget can be inferred from 
total mass estimates from dynamical modelling (e.g. Bell \& de Jong 2001; Cappellari et al. 2012) or gravitational lensing  (e.g. Treu et al. 2010), but 
this approach cannot unambiguously discriminate dark mass in cool dwarfs (bottom-heavy IMF) from that in stellar remnants (top-heavy IMF) or in 
non-baryonic dark matter.

An alternative method to constrain the mass locked in low-mass stars is to exploit stellar spectral features that depend strongly on surface gravity, and hence
distinguish giant from dwarf stars at similar effective temperature. In particular, red-optical and near-infrared spectral features that are strong in M-dwarfs and not 
present in cool giants (or vice versa) can potentially discriminate the low-mass stellar content  in unresolved galaxies. 
The Wing--Ford band (WFB) of FeH at 9915\,\AA\ (Wing \& Ford 1969) and the Na\,{\sc i} doublet (8190\,\AA) were identified as possible giant-to-dwarf 
star indicators and exploited in 
early works (e.g. Spinrad \& Taylor 1971;  Whitford 1977; Cohen 1978; Faber \& French 1980; Carter, Visvanathan \& Pickles 1986; Couture \& Hardy 1993). 
Some of these papers (e.g. Spinrad \& Taylor, Faber \& French, both based on Na{\,\sc i}) suggested a substantial fractional contribution of dwarf stars to the 
integrated light of nearby galaxies, while others (e.g. Whitford, based on the WFB) inferred a much lower dwarf-to-giant ratio. 
Similarly, Carter et al. reported tension between the two indicators for a sample of 14 galaxies, with Na{\,\sc i} requiring a larger fraction of dwarf light
than the WFB.

The work of the 1970s--1990s was hampered by the limitations of spectral synthesis models at that time. In particular, 
there were no large empirical infrared stellar libraries covering the full range of spectral types relevant to the problem. Moreover, 
knowledge of the range of elemental abundance variations in galaxies was not yet well constrained; nor was the machinery required to account for such 
variations through synthetic spectral modelling yet developed. Emphasis in the study of the IMF from resolved populations shifted somewhat to the 
Ca\,{\sc ii} triplet (which is strong in giants and weak in dwarfs), with Cenarro et al. (1993) concluding that either calcium was underabundant in giant ellipticals
or that a dwarf-enriched IMF was required. 

A new impetus was given to such studies by the publication of the IRTF spectral library by Rayner, Cushing \& Vacca (2009), which finally provided 
empirical library spectra for cool stars across the red/infrared regime. Building models from these stars to predict the integrated spectra of old populations,
and comparing them to observations of eight giant elliptical galaxies, van Dokkum \& Conroy (2010) argued that a strongly dwarf-enriched IMF was 
required to account for the strength of the Na{\,\sc i} and (to a lesser extend) the WFB. Their spectra could be fit only by adopting very bottom-heavy 
IMFs, e.g. a power law with exponent $x$\.=\,3 
(where ${\rm d}N(M)$\,$\propto$\,$M^{-x}{\rm d}M$), in which a majority of the mass is in $\la$0.15$\,M_\odot$ stars with 
$T_{\rm eff}\la3000$\,K.
The stellar mass-to-light ratio $M_*/L$ for the IMF favoured by van Dokkum \& Conroy is a factor of  3--4 times larger than the Chabrier-like IMF often assumed in 
extra-galactic contexts. Hence verifying this result, and determining the types of galaxies affected, has wide implications. 

Conroy \& van Dokkum (2012a, hereafter CvD12a) have since generalized their spectral synthesis models to make predictions for simple stellar populations 
(SSPs) with a range of IMFs, ages, metallicities and abundance ratio patterns.


Spurred in part by the original van Dokkum \& Conroy (2010) result, several groups have recently investigated the behaviour of IMF-sensitive indices in larger
galaxy samples. Spiniello et al. (2012) studied  the Na{\,\sc i} doublet in luminous red galaxies from the Sloan Digital Sky Survey (SDSS) with velocity dispersions 
$\sigma$=200--335\,\kms), finding this feature to increase with $\sigma$, requiring  an steepening IMF or an increase in Na/Fe 
enhancement (or both), with increasing galaxy mass.
Subsequently, Conroy \& van Dokkum have analysed new high-S/N spectroscopy including the WFB, the Na\,{\sc i} doublet and the Ca\,{\sc ii} triplet, as well as 
blue/optical features, for a sample of 38 nearby galaxies (van Dokkum \& Conroy 2012; Conroy \& van Dokkum 2012b, the latter CvD12b hereafter). 
In this work, they apply a sophisticated full-spectrum fitting 
method with many model components to describe abundance variations and possible confounding parameters. 
They conclude that dwarf-enrichment increases with velocity dispersion and (perhaps more strongly) with [Mg/Fe], ranging from Chabrier-like to 
heavier-than-Salpeter IMFs. Although all three of the ``classical'' indicators employed in this work seem to support dwarf-enriched IMFs in the most
massive ellipticals, the details depend on which spectral features are included in the fit. Perhaps to compensate for this tension, 
the models require enhancements of Na/Fe by up to an order of magnitude over the solar ratio. 
Most recently, Ferreras et al. (2012) used a very large sample of SDSS early-type galaxies to confirm the strong increase in Na{\,\sc i} with velocity dispersion
in the $\sigma$\,=\,100--300\,\kms\ range, interpreting the results as a steepening of the IMF slope. (They did not explicitly consider the effects of Na/Fe variations.)

For all but the most nearby galaxies, the Wing--Ford band is redshifted beyond the limits of current ``optical'' CCDs, and infrared detectors are required. 
In this paper we present new observations of the Wing--Ford band for red-sequence galaxies in the Coma cluster, using the infrared Fibre Multi-Object Spectrograph
(FMOS; Kimura et al. 2010) at the 8.2m Subaru telescope on Mauna Kea. Although galaxies at the distance of Coma are individually fainter than those 
studied by Conroy \& van Dokkum (2012b), we can exploit the dense concentration of galaxies in the cluster, and the multiplex capability of FMOS, to observe dozens of galaxies 
simultaneously. Because the Coma galaxies have a range of radial velocities due to their motions within the cluster potential, stacking their spectra in the 
rest-frame allows rejection of pixels contaminated by OH sky-line residuals and other artefacts fixed in the observed wavelength frame. 
Our galaxy sample spans a range in velocity dispersion and other properties, derived in previous work from optical spectroscopy (Price et al. 2011; Smith et al. 2012),
so that we can explore possible variations in the IMF with galaxy mass and control for element abundance effects. Besides the WFB, our spectral range also
includes a Ca{\,\sc i} line which we exploit as an IMF indicator for the first time.

The outline of this paper is as follows. Section~\ref{sec:obs} describes the observations and data reduction. The construction of composite spectra
is detailed in Section~\ref{sec:compo}. The results are presented in Section~\ref{sec:results}, starting with the global composite spectrum 
(Section~\ref{sec:globalcomp}), the measurement of spectral indices and their comparison to model predictions (Section~\ref{sec:indices}), and then
dividing the sample into four bins in velocity dispersion (Section~\ref{sec:sigmabin}) and Mg/Fe ratio (Section~\ref{sec:mgfebin}), to explore the 
dependence of the IMF on mass and star-formation timescale. In Section~\ref{sec:discuss}, we discuss
our results in the context of relevant recent work, and summarize our conclusions in Section~\ref{sec:concs}.


\begin{figure*}
\includegraphics[angle=270,width=178mm]{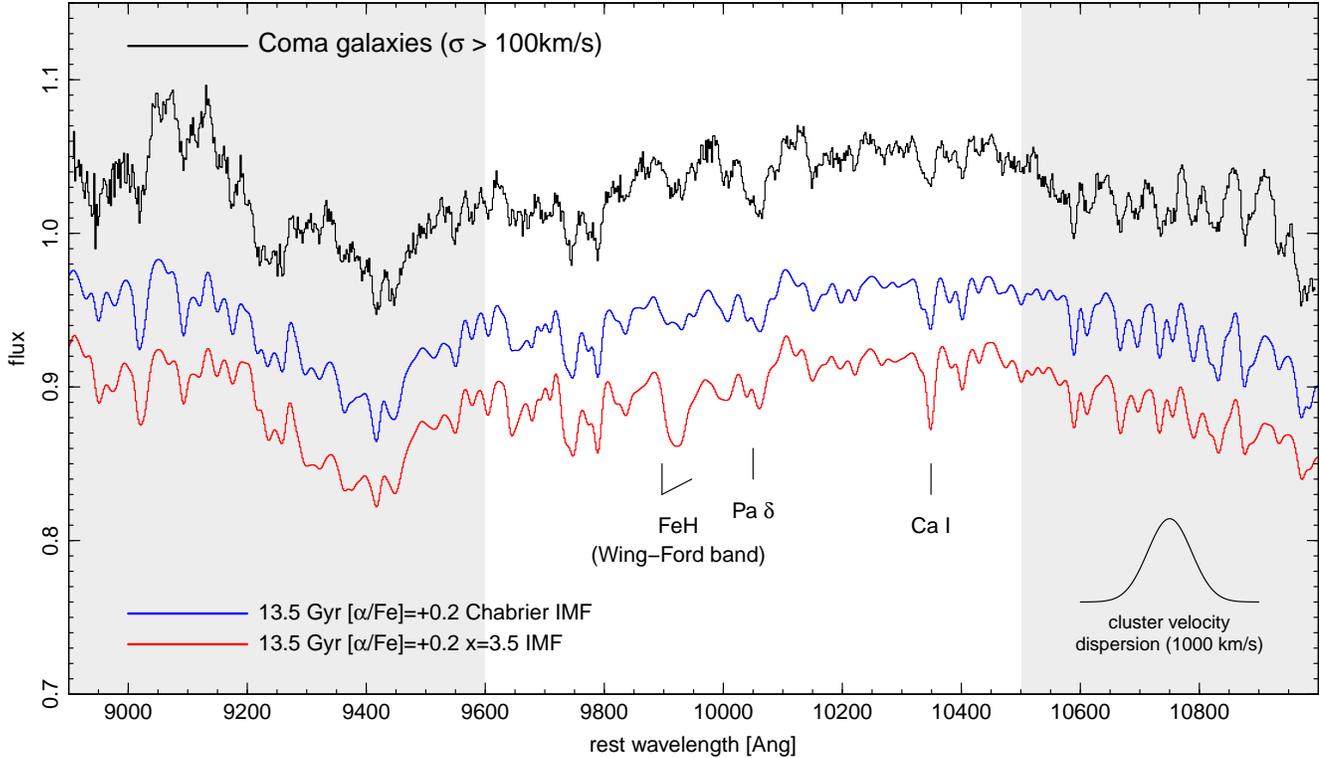}
\caption{Composite spectrum constructed from the 59 observed galaxies with $\sigma$\,$>$\,100\,\kms\ (top, black). The lower
spectra show SSP models from CvD12a, with different IMFs, smoothed to match the effective velocity dispersion of the stacked data. 
The $x$\,=\,3.5 IMF is extremely bottom-heavy and is plotted here to emphasize the IMF dependence of the WFB
and Ca{\,\sc i} line. 
Subsequent figures show only on the rest-frame 9600--10500\,\AA\ region (white background) where the atmospheric transmission is clean and which includes the
IMF-sensitive spectral features. For reference, we show a gaussian profile representing the shifts induced by the velocity dispersion within the cluster; 
any residual sky-subtraction artefacts are effectively smoothed over this scale in the composite spectrum.}
\label{fig:compo_fullwav}
\end{figure*}

\section{Observations and data reduction}\label{sec:obs}

We observed Coma Cluster galaxies with FMOS on the nights 2012 May 6--7, using 
high spectral resolution mode  in ``J-short'' configuration. This setting provides spectral coverage over the range 9130--11270\,\AA\
(corresponding to 8920--11010\,\AA\ at the average redshift of Coma), 
with resolution $4.6$\,\AA\ FWHM, sampled at 1.25\,\AA\ per pixel. 
FMOS features two separate (and somewhat different\footnote{see Kimura et al. (2010)}) spectrographs, IRS1 and IRS2, each receiving half of the 400 total fibres. 
The fibres project to 1.2\,arcsec in diameter, corresponding to 0.6\,kpc\ at our adopted distance for Coma (100\,Mpc, i.e. $H_0=72$\,\kms\,Mpc$^{-1}$).

Targets were selected from the sample of 362 Coma galaxies analysed by Smith et al. (2012), which spans a range from the most massive ellipticals 
($r_{\rm petro}$\,$\approx$\,$12$, $M_*$\,$\approx$\,$10^{12}\,M_\odot$, $\sigma\approx300$\,\kms) down to the dwarf galaxy population 
($r_{\rm petro}$\,$\approx$\,$18$, $M_*$\,$\approx$\,$10^{9}\,M_\odot$, $\sigma\approx30$\,\kms).
The Smith et al. sample excludes galaxies with significant ongoing star formation using a cut on H$\alpha$ equivalent width. 
FMOS deploys fibres within a 30-arcmin diameter field of view; however each fibre can patrol only a small part of the field, so that the  
ability to pack fibres closely is quite limited. To maximise the number of galaxies observed, we used four fibre configurations with pointings offset by a few arcminutes
and different field rotations (constrained by the availability of guide stars). 
Most galaxies in the sample were observed in more than one configuration, through different fibres, and in many cases with both spectrographs.  
In total, 121 galaxies were observed at least once. However, the faintest galaxies contribute little signal even if stacked.
For the analysis in this paper, we restrict our attention to 92 target galaxies having $\sigma$\,$>$\,50\,\kms.

The observations were made in cross-beam-switching mode, in which two fibres are allocated to each target and the telescope nodded (usually every 900\,seconds)
to move the galaxies from one fibre to the other between integrations. 
This allows the sky background for each target to be determined using the same fibre and detector pixels as for the galaxy light,
while also keeping each target within a fibre for all of the integrations (equivalent to nodding along the slit in single-object spectroscopy). 
The fibres in each pair are separated by 90 arcsec which is sufficiently large to avoid 
contamination of the sky by the outer part of the target galaxy for all objects in our sample. 
The combined integration times for the four fibre configurations were 2.33, 1.50, 2.00 and 2.00 hours, 
for a total of 7.83 hours total integration. Observing conditions were fairly good, with seeing 0.6--1.1\,arcsec 
and occasional  passing cirrus on the second night.

Initial data reduction was performed using the standard FMOS {\sc fibre-pac} pipeline (Iwamuro et al. 2012), yielding extracted, calibrated one-dimensional 
spectra for each fibre. 
At the redshift of Coma, the WFB is observable in a region of very clean atmospheric transmission.
Typical telluric absorption features in this window are less than 1\,per cent in amplitude. The  {\sc fibre-pac} pipeline attempts to correct for 
atmospheric absorption using stars observed simultaneously with the science targets, but this requires correcting for the intrinsic spectral 
signatures of those stars, in particular the Pa\,$\delta$ line, lying $\sim$150\,\AA\ short-wards of the redshifted WFB. 
Despite taking particular care to improve the removal of this feature by manually tuning the spectral type assumed for the star, 
we were not able to remove these artefacts to our satisfaction. In practice we opted instead to mask the affected region when stacking the 
spectra.

\begin{figure*}
\includegraphics[angle=270,width=178mm]{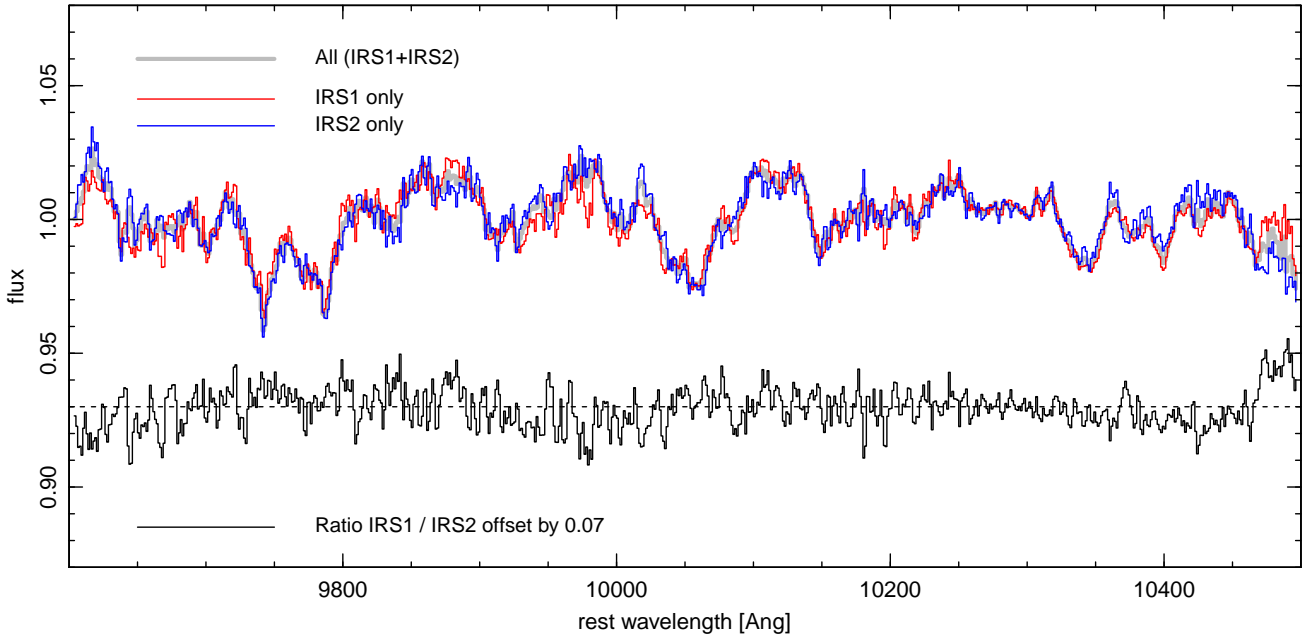}
\caption{Comparison between composite spectra computed using each of the two FMOS spectrographs, IRS1 and IRS2.  
We have combined all $\sigma$\,$>$\,100\,\kms\ galaxies observed on each spectrograph; the two sets of raw data 
are independent, though some {\it galaxies} contribute to both stacks.}
\label{fig:sgcompare}
\end{figure*}

\section{Composite spectra}\label{sec:compo}

The results of this paper are based upon analysis of composite spectra created by stacking the spectra of many galaxies, after masking ``bad'' 
intervals in observed wavelength and shifting to the rest frame. Because the sample spans a significant range in redshift due to the large virial motions
within the cluster (equivalent to FWHM\,$\sim$\,80\,\AA), this process fills the ``gaps'' imposed by residual sky contamination and other artefacts which are fixed in the 
observed wavelength frame. 

FMOS features an OH suppression mask to suppress the wavelength regions affected by strong airglow lines. However, these regions must
still be masked in software, since they otherwise appear as low-flux notches within the extracted spectra. The masks in the two spectrographs are
different, with IRS1 masking many more weak lines than IRS2. Moreover, during reduction of our data, we discovered that the mask in IRS2 was
misaligned causing masking to occur at wavelengths $\sim$60\,\AA\ shortwards from the target sky lines. 

We impose software masking at the locations of the OH-suppression notches (in the case of IRS2, at their actual location, not at the position of the sky lines
they were intended to mask, which are adequately subtracted in the reduced spectra). 
Additionally we mask a 20\,\AA\ region centred on (un-redshifted) Pa\,$\delta$, to remove any residuals arising from poor
atmospheric absorption corrections. After masking, we correct to the rest frame and re-bin to a common wavelength scale of 9600--10500\,\AA, sampled at 
1.25\,\AA\ intervals (the native sampling of the original spectra). These spectra, of which there are up to four per galaxy, form the input to the compositing procedure. 

To combine the spectra, we first apply a simple median over the inputs to obtain an initial estimate for the spectral shape. 
Each spectrum is then matched to this median using a 4th-order polynomial correction to remove residual throughput variations. 
After this we identify and reject outlying pixels (beyond $\pm$5$\sigma$ from the average over all input spectra), and finally combine
surviving pixels using a mean, weighted by the inverse of the variance spectrum. (For the latter, we use the background noise spectrum 
estimate from {\sc fibre-pac} together with a component representing poisson noise in the object spectra.)

Errors in the composite spectra were estimated both by using the formal error on the weighted mean, and by computing the composite using 
100 bootstrap-resampled realisations of the set of input spectra. The two error estimates agree to within 5\,per cent in the region of the WFB and within 
20\,per cent across the whole 
wavelength range. However, the rebinning of the spectra onto a common rest-frame wavelength scale inevitably introduces correlations in the errors between 
adjacent pixels. In our analysis we will generally make use of the bootstrap realisations of the composite spectrum to propagate the uncertainties appropriately.

\begin{figure*}
\includegraphics[angle=270,width=175mm]{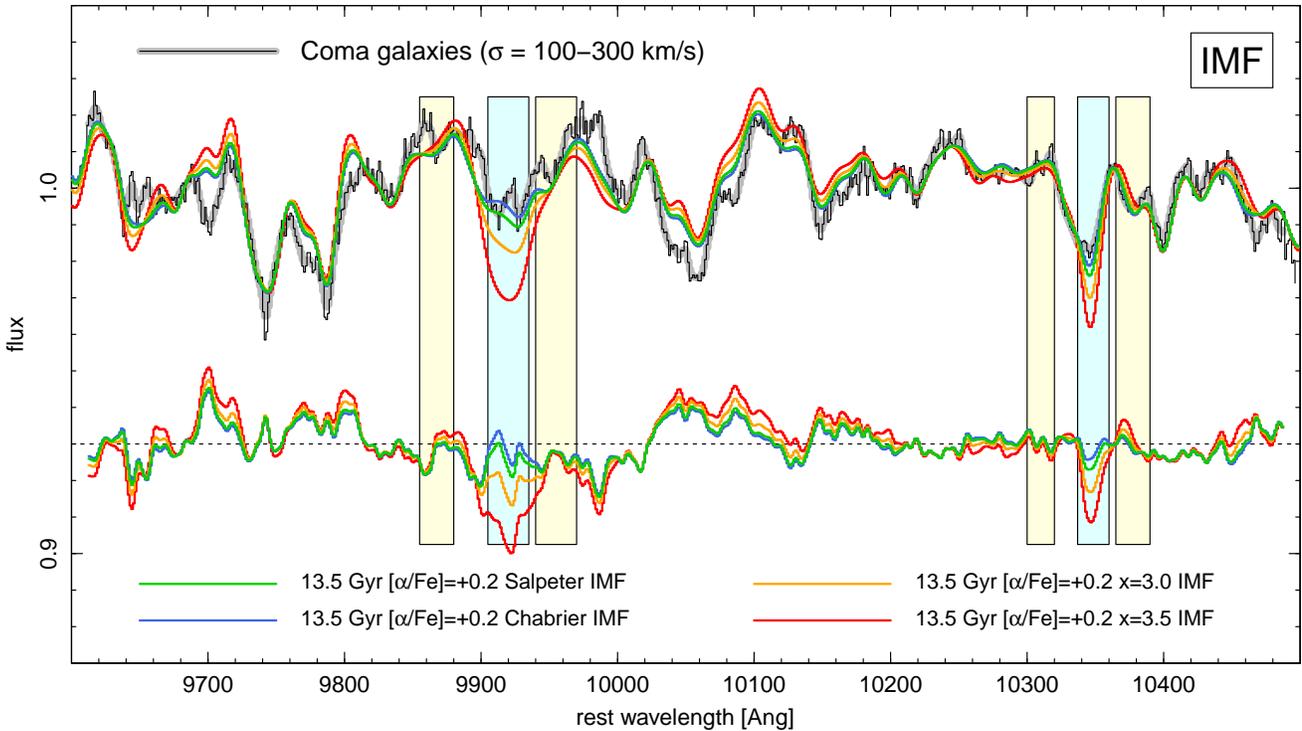}
\caption{Composite spectrum constructed from the 59 observed galaxies with $\sigma$\,$>$\,100\,\kms\ (thin black line). The thick grey line shows  the same spectrum smoothed
by an amount corresponding to $\sigma\approx$\,100\,\kms\ to suppress small-scale noise while retaining the intrinsic resolution of the spectrum. 
The coloured lines are 13.5\,Gyr models from CvD12a, with slightly enhanced $\alpha$/Fe ratio (matching the optically-determined Mg/Fe), and a range of different IMFs. 
The lower section shows the flux ratio (model/data) for the same set of models, offset by 0.07.
The vertical bands show the feature and pseudo-continuum band-passes defining the FeH0.99 index of CvD12a and our newly-defined CaI1.03 index.
Note that the red pseudo-continuum of FeH0.99 is higher in the data than the models; hence although the ``floor'' of the absorption matches the Chabrier 
IMF model, the measured index strength is closer to that of the Salpeter model (see Sec~\ref{sec:indices}).
}
\label{fig:compo_biggx_imf}
\end{figure*}

\begin{figure*}
\includegraphics[angle=270,width=175mm]{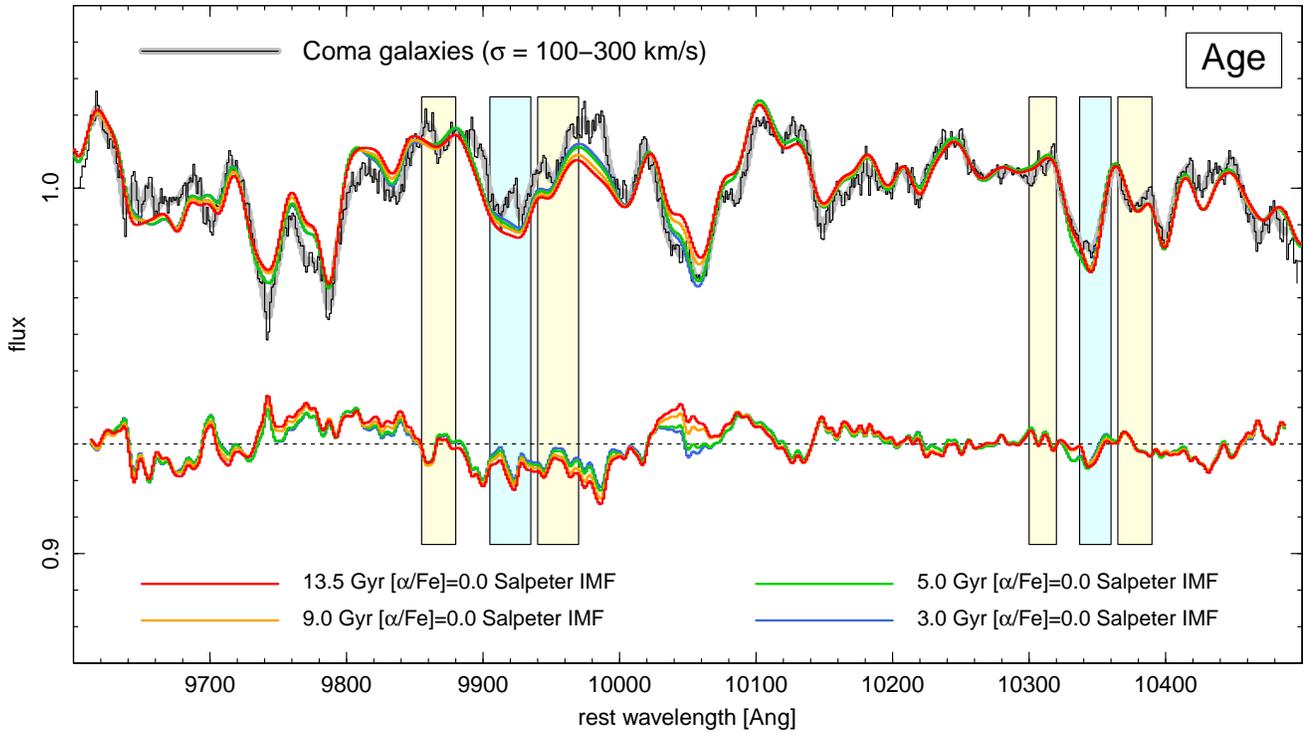}
\caption{Data as in Figure~\ref{fig:compo_biggx_imf}, but compared to models for  a range of different ages, 
all having solar $\alpha$/Fe ratio and Salpeter IMF.
The figure demonstrates the robustness of the models against age variations (except around Pa\,$\delta$ at
$\sim$10050\,\AA). Note that none of these models has $\alpha$/Fe appropriate to the observed galaxy sample.
}
\label{fig:compo_biggx_age}
\end{figure*}

\begin{figure*}
\includegraphics[angle=270,width=175mm]{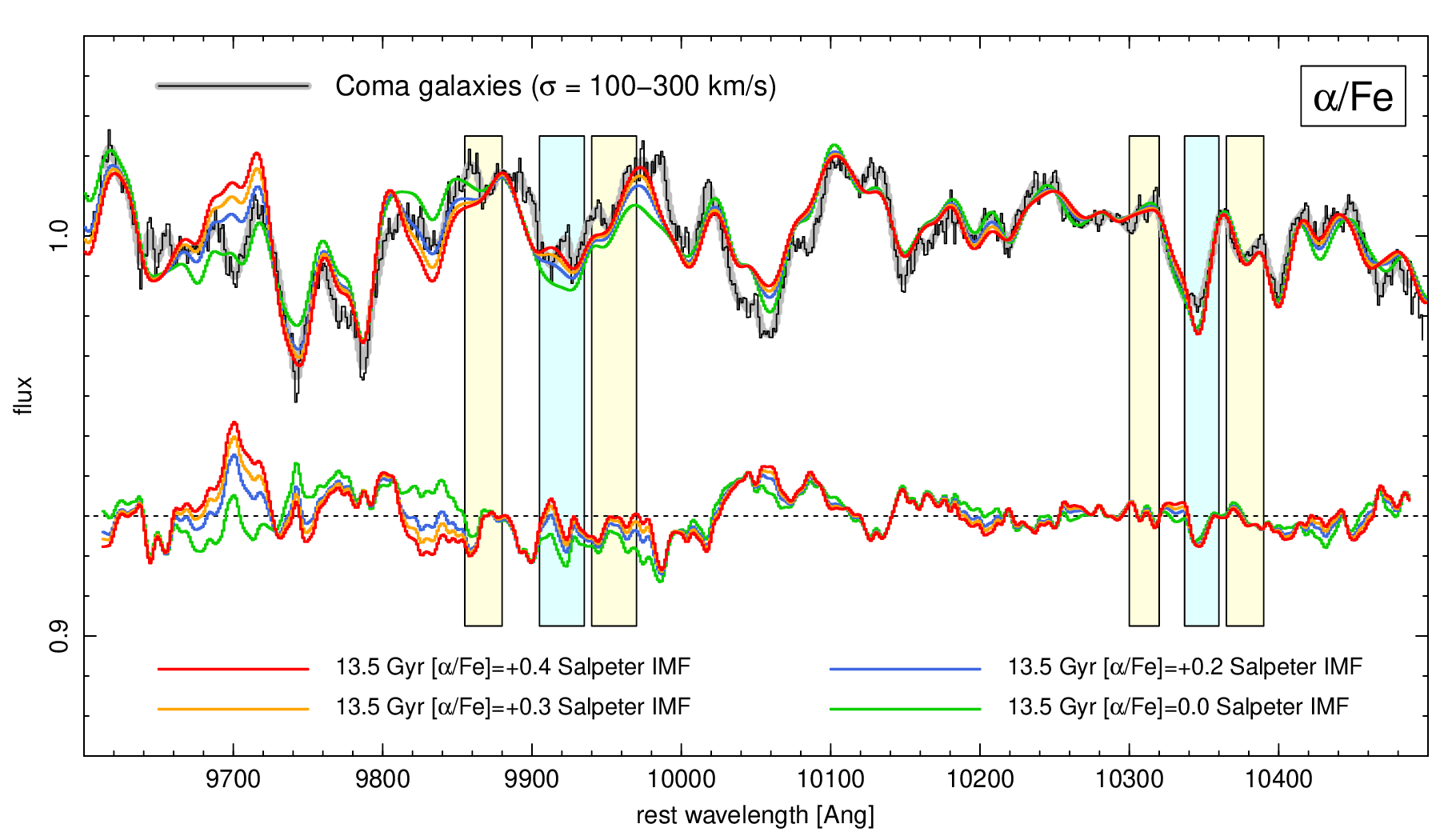}
\caption{Data as in Figure~\ref{fig:compo_biggx_imf}, but compared to models for  a range of different $\alpha$/Fe ratios,
all having 13\,Gyr age and Salpeter IMF.}
\label{fig:compo_biggx_afe}
\end{figure*}

In construting the stacked the spectra, we combine galaxies with a range of different velocity dispersions. Broadening
all the spectra to the largest velocity width is not appropriate for these data, because it would spread 
systematic errors (e.g. from sky-subtraction residuals) and missing data into most of the pixels in each
spectrum. In our subsequent analysis, we account for the effects of velocity broadening by convolving 
model spectra to match the weighted mean velocity dispersion of galaxies in the stacks\footnote{We 
have confirmed all results using a convolution kernel taking which takes account of 
the range of velocity dispersions entering the stack, with their appropriate weights. All measured 
line-strength indices are insensitive to which method is used, to within one tenth of the quoted measurement 
error. We quote results from the simpler scheme throughout the paper.}.




\begin{figure*}
\includegraphics[angle=270,width=178mm]{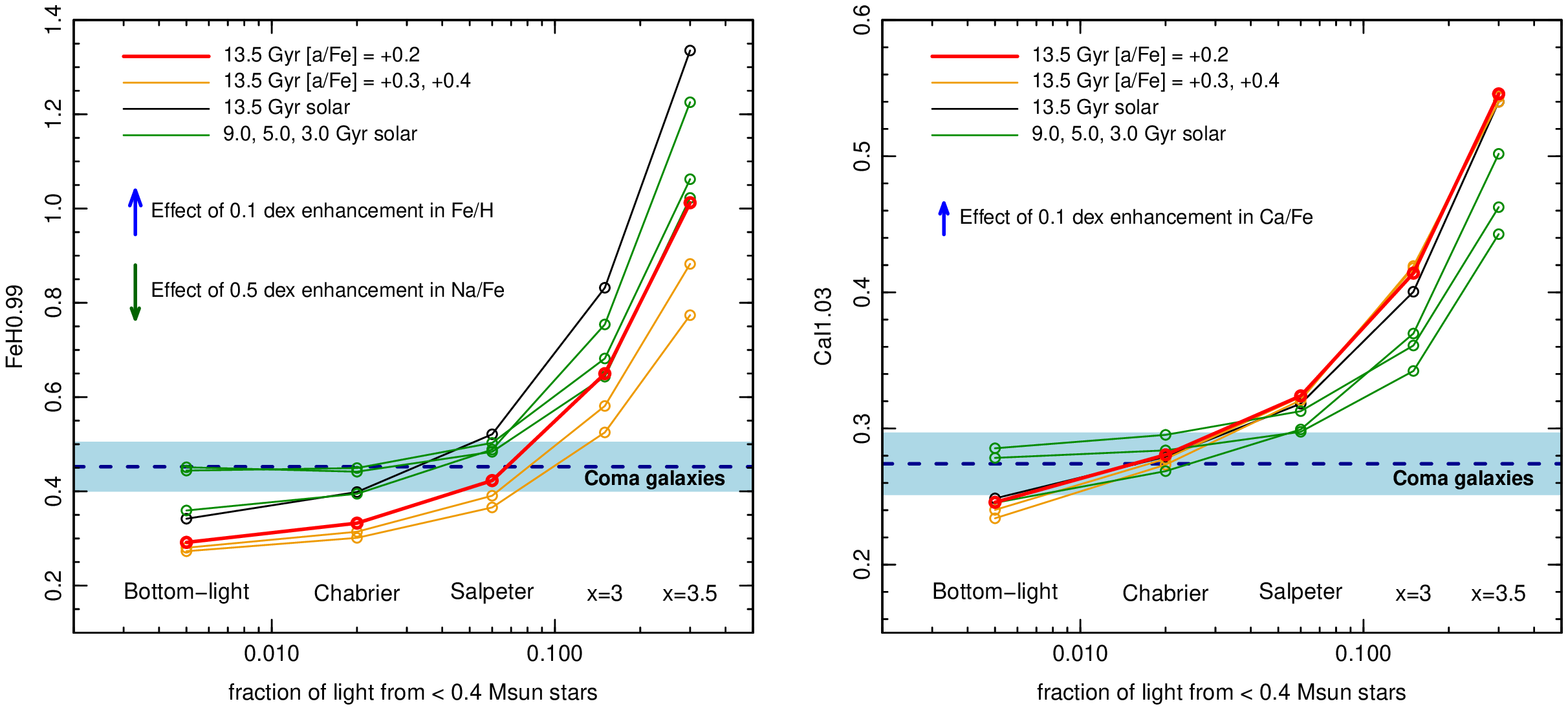}
\caption{Absorption indices measured from the global composite (dashed line and blue-shaded 1$\sigma$ error interval), 
compared to indices measured on the CvD12a model spectra. The Wing--Ford band index FeH0.99 is defined as in CvD12a, 
while CaI1.03 is newly defined in this paper (see text).
The indices are measured on the models after smoothing to the  mean velocity dispersion appropriate to the data, and 
continuum matching with a low-order polynomial correction. Vectors indicate how the model predictions for FeH0.99 depend on 
Fe/H and Na/Fe, and how CaI1.03 is affected by Ca/Fe. 
}
\label{fig:biggx_indices}
\end{figure*}

\section{Results}\label{sec:results}

\begin{figure*}
\includegraphics[angle=270,width=178mm]{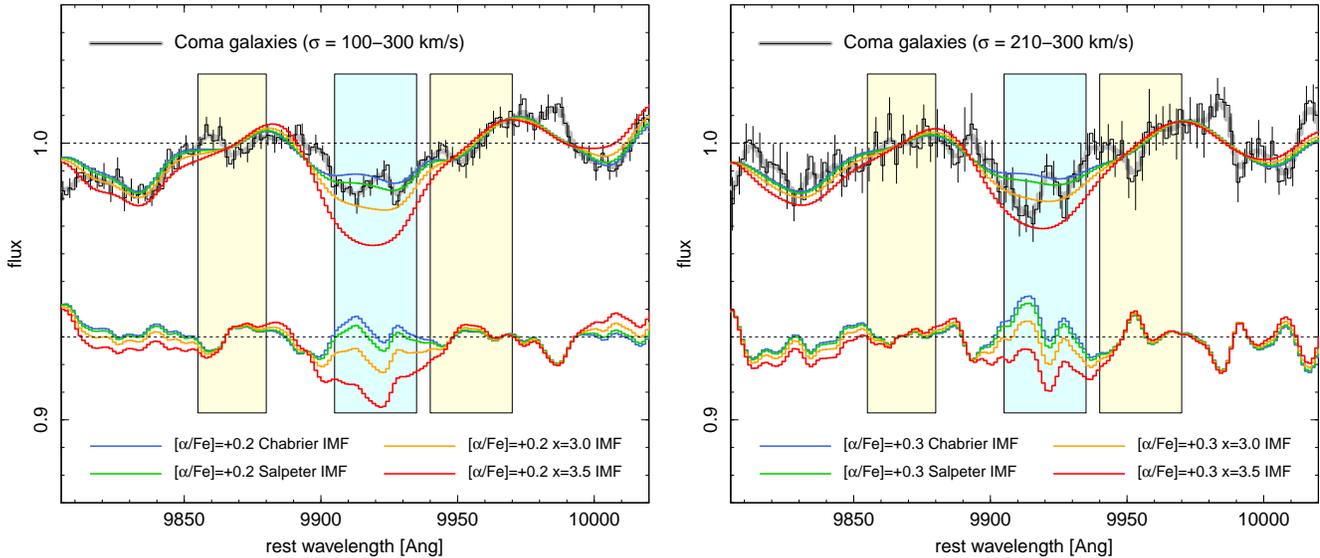}
\caption{The WFB region of the composite spectra for the global composite and for the highest $\sigma$ bin. In this figure, the spectra have been 
normalised to unity at the FeH099 pseudo-continuum bands. Bootstrap-derived error bars are shown on every second pixel.}
\label{fig:indexspec}
\end{figure*}

\begin{figure*}
\includegraphics[angle=270,width=178mm]{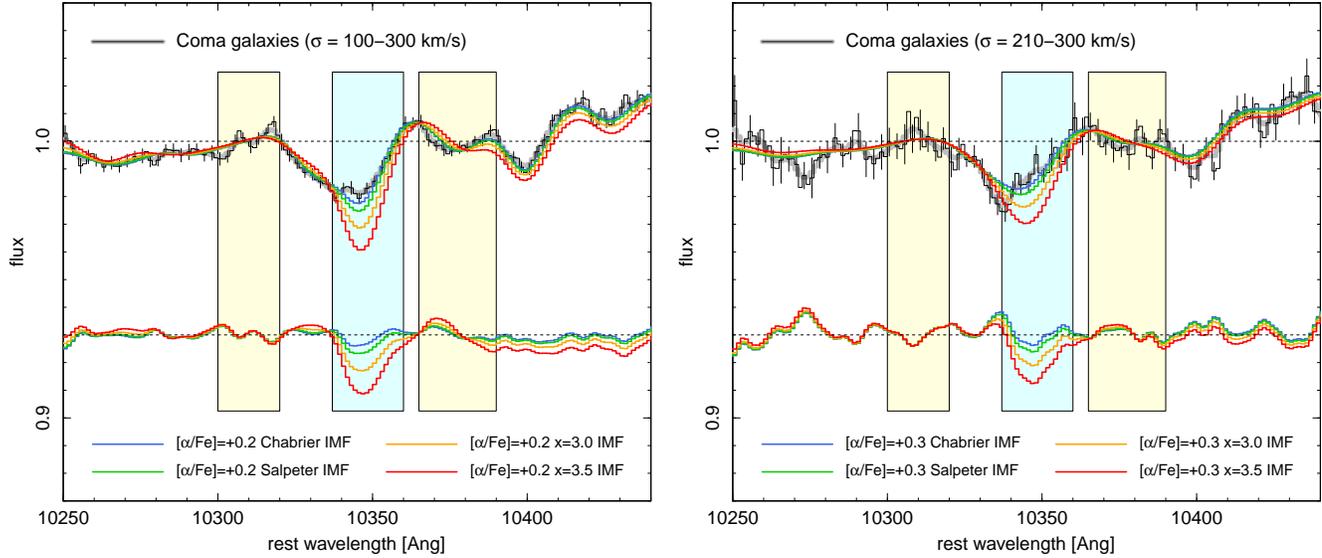}
\caption{As for Figure~\ref{fig:indexspec}, but for the CaI1.03 index.}
\label{fig:indexspec_cai}
\end{figure*}

\begin{figure*}
\includegraphics[angle=270,width=178mm]{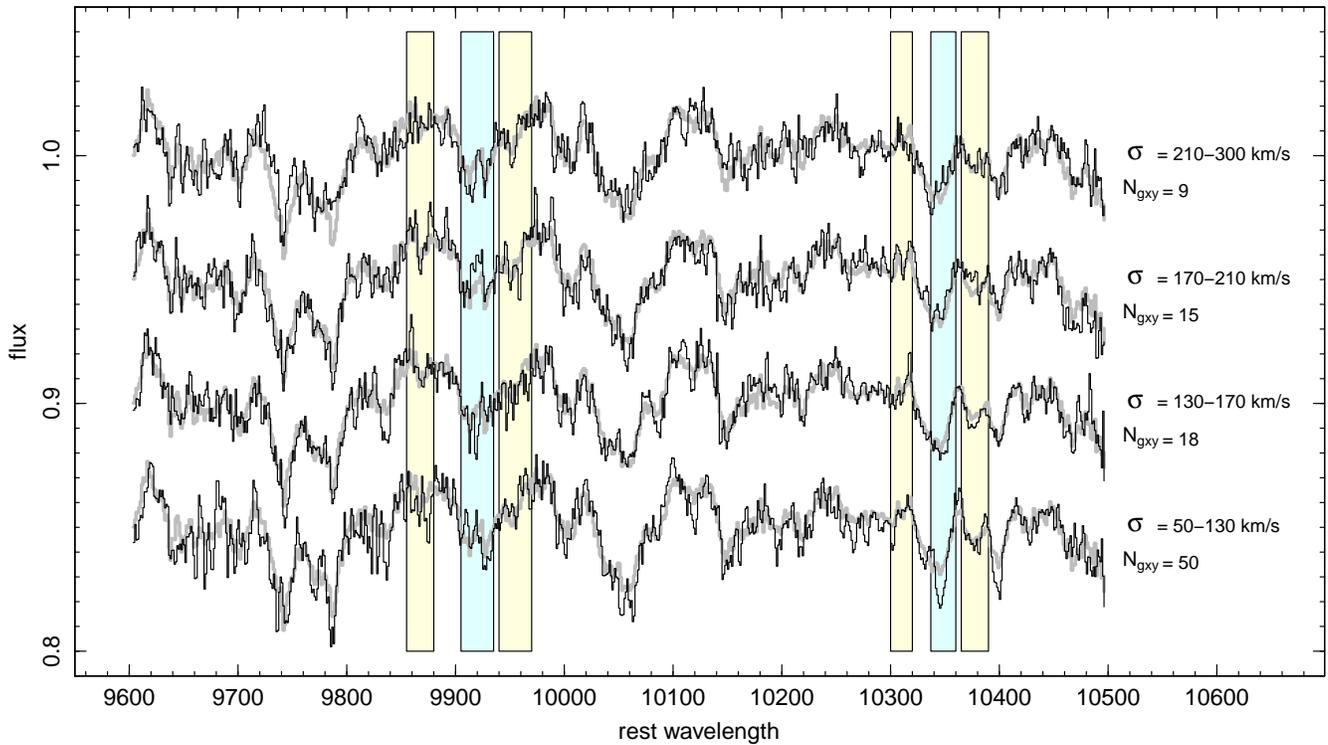}
\caption{Composite spectra computed in four bins of velocity dispersion (black lines). In each case, the thick grey line shows the global
composite of $\sigma$\,$>$\,100\,\kms\ galaxies for comparison. In this figure, no smoothing has been applied to either the $\sigma$-split or the 
global composite.}
\label{fig:compo_sigsplits}
\end{figure*}

%

\begin{figure*}
\includegraphics[angle=270,width=178mm]{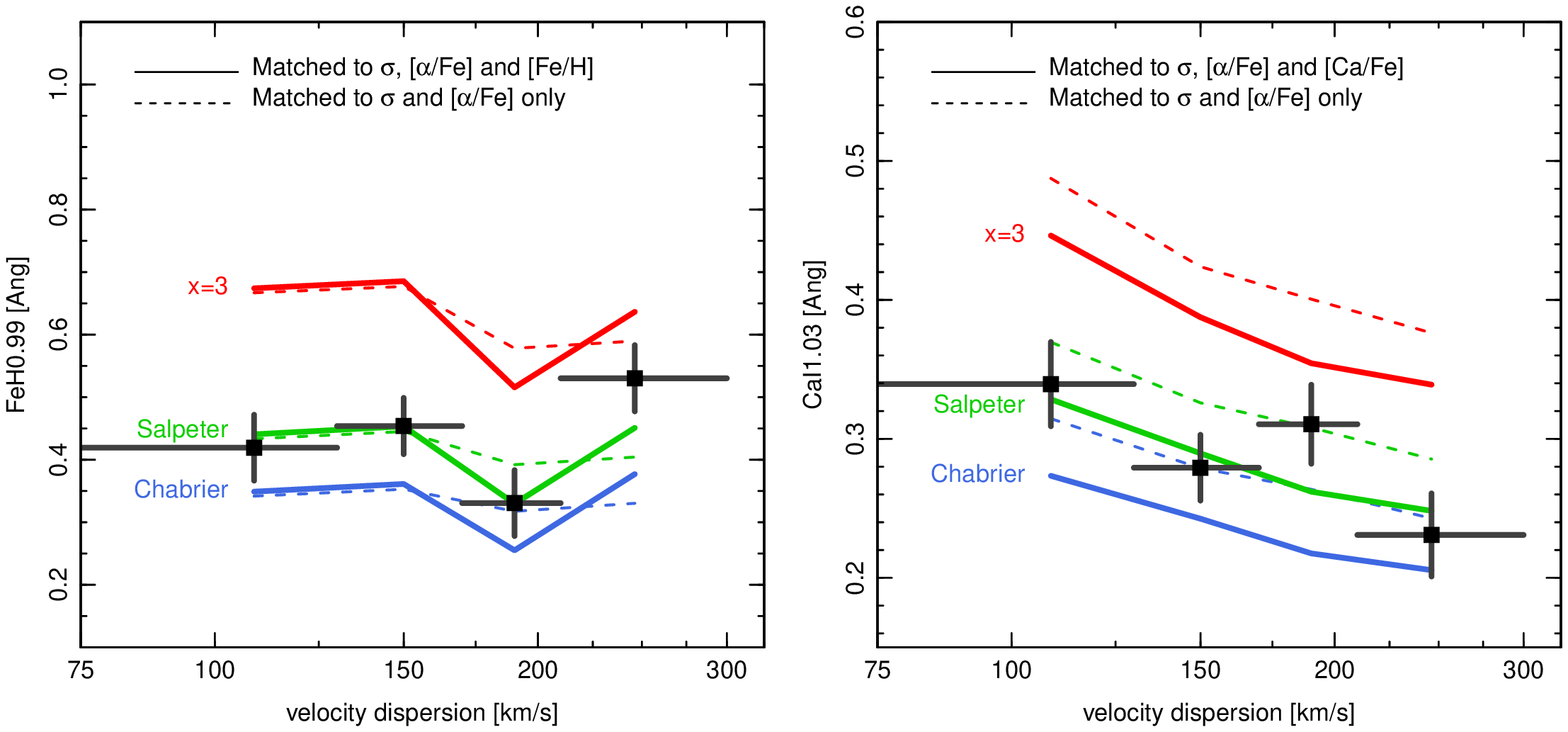}
\caption{Variation in the FeH0.99 and CaI1.03 indices as a function of velocity dispersion. Bins in $\sigma$ are as in Figure~\ref{fig:compo_sigsplits}.
The coloured lines show the same indices measured from constant-IMF CvD12a models tuned to match the properties of each bin.  
The models are smoothed to the contribution-weighted mean velocity dispersion, interpolating between models of different $\alpha$/Fe
to match the (similarly averaged) Mg/Fe. In the left-hand panel, the solid lines include a correction to the measured Fe/H (the dashed lines assume solar Fe/H). 
In the right-hand panel, Fe/H is assumed solar (the CaI1.03 index is insensitive to the iron abundance). 
Here, the dashed lines assume calcium behaves as a generic $\alpha$ element, i.e. Ca/Fe=$\alpha$/Fe=Mg/Fe. 
The solid line includes a correction to Ca/Fe as derived from the optical spectra; applying this correction changes the preferred
IMF from Chabrier to Salpeter.
}
\label{fig:indices_sigsplits}
\end{figure*}

\subsection{Global massive-galaxy composite}\label{sec:globalcomp}

We begin by examining qualitatively the spectrum created by stacking the 59 galaxies with $\sigma$\,$>$\,100\,\kms. 
Within this sample, the mean velocity dispersion, weighted according to the contribution of each galaxy to the stack, is 180\,\kms.
The velocity dispersion range $\sigma$\,=\,125--250\,\kms\ contributes 90 per cent of the total weight.
Using line-strength indices in the optical range (4000--5500\,\AA), Smith et el. (2012) derived metallicities and abundance estimates 
through comparison with the SSP models of Schiavon (2007). The average optically-derived parameters, again weighted by contribution
to the stacked FMOS spectrum are: [Fe/H]=$0.00$, [Mg/Fe]=$+0.25$ and SSP-equivalent age 10\,Gyr. 

Figure~\ref{fig:compo_fullwav} shows the composite spectrum over the full spectral range observed.
The total signal-to-noise ratio of this spectrum is $\sim$170 per angstrom.
A visual comparison with the SSP models of CvD12a
shows that the observed features correspond closely to the expected absorption signatures of old stellar populations. 
We restrict all further analysis to the 9600--10500\,\AA\ spectral range which includes the IMF-sensitive WFB and Ca{\,\sc i} line.

In Figure~\ref{fig:sgcompare} we divide the same set of input data into two, computing composite spectra for each of the FMOS spectrographs separately. 
Comparison between the IRS1 and IRS2 composites shows that despite their different characteristics (in particular the much heavier OH suppression mask
in IRS1), there is excellent agreement between the average spectra obtained from the two spectrographs.

In Figure~\ref{fig:compo_biggx_imf}, we compare the observed composite spectrum against SSP models from CvD12a, with 
old ages (13.5\,Gyr), solar Fe/H and mildly-enhanced $\alpha$/Fe ratio. The publicly-available CvD12a models cover either
enhanced $\alpha$/Fe ratio at 13.5\,Gyr or younger ages at solar abundance ratios. We choose in this figure to match the 
enhanced $\alpha$/Fe ratio rather than the slightly younger ages which
may be relevant to our sample. (Note that for realistic, non-SSP star-formation histories, the ages derived from optical spectra will be more
affected by recent episodes of star formation, while the infrared spectra analysed here will be dominated by the oldest populations.)
Each model has been broadened the weighted mean velocity dispersion for the stack, and matched to the overall shape of the 
observed spectrum using a fourth-order polynomial correction. 
The smoothing scale has been verified by a direct $\chi^2$-fitting of the models to the observed spectrum; in practice, we prefer to fix this to the
average velocity dispersion which is securely measured from optical data. 

Qualitatively, there is very good agreement between observed and model spectra in regions that are not sensitive
to the IMF, with all the main spectral features well reproduced. A systematic discrepancy is seen in the feature at $\sim$10050\,\AA, 
which includes the Pa\,$\delta$ line as well as TiO absorption, with stronger absorption observed than in the models, causing residuals at the
$\sim$1 per cent level over a $\sim$50\,\AA\ interval. (A very similar discrepancy is also evident in Figure~1 of Conroy \& van Dokkum 2012b.)  
More localised 1 per cent discrepancies are also seen at $\sim$9700--9800\,\AA.

In the WFB region, the bottom of the absorption feature lies close to the models with Chabrier or Salpeter IMF.
However, it should be noted that the spectral region immediately red-wards of the WFB is poorly matched by all models. 
Hence, inferring the IMF through this comparison is somewhat degenerate with the treatment of the continuum 
matching or, in the case of spectral indices, with the placement of the pseudo-continuum band-passes. 
Furthermore, the detailed shape of the absorption band itself does not agree well with any of the CvD12a models: the observed spectrum
shows a bump at 9920\,\AA, while the models at this resolution show a flat-bottomed absorption. This discrepancy does not seem to
arise from a resolution mismatch; reducing the assumed velocity dispersion to introduce more structure in the WFB results in an unacceptable
fit elsewhere in the spectrum, and would be incompatible with the optically-measured $\sigma$.

At 10345\,\AA\ there is another absorption feature which is sensitive to the IMF, according to the models. 
This feature is marked, as Ca{\,\sc i}, among other possible IMF indicators, in Figure 10 of CvD12a, but was not pursued further in that paper 
nor (to our knowledge) in any other previous work. The Ca{\,\sc i} line is in the red wing of a blend, with lines of Fe{\,\sc i} and Sr{\,\sc ii} 
contributing to the blue wing (identified from the Arcturus atlas of Hinkle, Wallace \& Livingston 1995). 
The observed spectrum shows weak absorption at Ca{\,\sc i},  apparently compatible with the Chabrier/Salpeter IMF, 
and inconsistent with the dwarf-enriched models. The immediate environment of  this feature is well fit by the models.


According to the CvD12a models, the 1$\mu$m spectra depend only weakly on age and element abundances, as we show in 
Figures~\ref{fig:compo_biggx_age}--\ref{fig:compo_biggx_afe}. Figure~\ref{fig:compo_biggx_age} compares the same observed composite spectrum 
against CvD12a models of varying age, all having the Chabrier IMF and solar $\alpha$/Fe. Relative to the oldest model, younger ages reduce the 
systematic discrepancies, but the improvement is substantial only in the 
$\sim$10050\,\AA\ region (i.e. near Pa\,$\delta$), and ages as low as 3--5\,Gyr would be required to match the strong absorption here. 
Since the mean optical SSP-equivalent age is $\sim$10\,Gyr, and the equivalent age in the infrared should be older than in the 
optical\footnote{In principle, large dust-obscured young sub-populations could violate this expectation in some galaxies, 
but this seems unlikely to affect our sample {\it on average}.},
this does not seem to be a tenable solution, and may point to some incompleteness in the model.
%
%
Figure~\ref{fig:compo_biggx_afe} similarly compares the observed composite spectrum against models with varying $\alpha$/Fe abundance ratio.
Increasing $\alpha$/Fe has strengthens the absorption at 9830\,\AA, improving the agreement immediately blue-wards of the WFB,
but to leads to weaker the absorption at  $\sim$9700\,\AA, increasing the discrepancy with the data here.

\begin{table*}
\caption{IMF-sensitive line strength indices measured from stacked spectra after binning as a function of velocity dispersion and Mg/Fe ratio. 
The first column indicates the parameter used to define the bin, while min and max are the limits for that parameter.  $N_{\rm gxy}$ is the number of 
galaxies in the stack. Velocity dispersion $\sigma$ is in units of \kms. We indicate the average SSP-equivalent stellar population parameters
for each bin from the optical spectroscopy of Smith et al. (2012). These averages are weighted according to the contribution of each galaxy to the 
stacked spectrum.}
\label{tab:sigsplit}
\begin{tabular}{lcccccccccc}
\hline
Bin & min & max & $N_{\rm gxy}$ & $\langle\sigma$[\kms]$\rangle$  & $\langle$[Fe/H]$\rangle$ & $\langle$[Mg/Fe]$\rangle$ & $\langle$[Ca/Fe]$\rangle$ & $\langle$Age[Gyr]$\rangle$ &
FeH0.99 [\AA] & CaI1.03 [\AA] \\
\hline
$\sigma\ [$\kms$]$ &100 & 300 & 59 & 180 & +0.00 & +0.25 & +0.07 & 10.2 & $0.452\pm0.055$ & $0.274\pm0.024$  \\
\hline
$\sigma\ [$\kms$]$ & \phantom{0}50 & 130 & 50 & 109 & $+0.01$ & $+0.20$ & $+0.03$ & 7.1 & $0.419\pm0.053$ & $0.339\pm0.030$  \\
$\sigma\ [$\kms$]$ & 130 & 170 & 18 & 150 & $+0.01$ & $+0.19$ & $+0.04$ & 9.7 & $0.454\pm0.045$ & $0.279\pm0.024$  \\
$\sigma\ [$\kms$]$ & 170 & 210 & 15 & 190 & $-0.07$ & $+0.31$ & $+0.10$ & 12.1 & $0.331\pm0.053$ & $0.311\pm0.028$  \\
$\sigma\ [$\kms$]$ & 210 & 300 &  \phantom{0}9 & 246 & $+0.05$ & $+0.27$ & $+0.08$ & 10.7 & $0.530\pm0.053$ & $0.231\pm0.030$  \\
\hline
$[$Mg/Fe$]$ & $-0.05$ & $+0.15$ & 34 & 114 & $+0.08$ & $+0.12$ & $-0.03$ & 6.9 & $0.391\pm0.061$ & $0.267\pm0.032$  \\
$[$Mg/Fe$]$ & $+0.15$ & $+0.22$ & 21 & 168 & $+0.04$ & $+0.19$ & $+0.05$ & 8.8 & $0.453\pm0.044$ & $0.314\pm0.021$  \\
$[$Mg/Fe$]$ & $+0.22$ & $+0.30$ & 16 & 188 & $+0.01$ & $+0.27$ & $+0.07$ & 9.4 & $0.529\pm0.056$ & $0.259\pm0.029$  \\
$[$Mg/Fe$]$ & $+0.30$ & $+0.46$ & 21 & 201 & $-0.12$ & $+0.36$ & $+0.13$ & 13.7 & $0.331\pm0.049$ & $0.300\pm0.027$  \\
\hline
\end{tabular}
\end{table*}

\subsection{Absorption indices and inferred IMF}\label{sec:indices}

To quantify our results, we compute ``Lick-style'' line-strength indices for the WFB and Ca{\,\sc i} features. For the WFB, we adopt the index 
defined as FeH0.99 in CvD12a:
feature band at 9905--9935\,\AA\ and pseudo-continua at 9855--9880\,\AA\ and 9940--9970\,\AA. To measure Ca{\,\sc i}, we define a 
new index (CaI1.03) with feature band at 10337--10360\,\AA\ and pseudo-continua at 10300--10320\,\AA\ and 10365--10390\,\AA. 
The feature band is carefully placed to isolate the IMF-sensitive Ca\,{\sc i} line, rather than the blended neighbouring absorption, and to minimise
dependence on age and $\alpha$/Fe ratio, through examination of Figures~\ref{fig:compo_biggx_imf}--\ref{fig:compo_biggx_afe}.
When measuring the indices on the CvD12a models, we first match these models to the observed spectrum as before, using the 
weighted mean velocity dispersion of 
the galaxy stack, and allowing a fourth-order continuum correction to be fit independently for each model. The results are insensitive
to the order of the correction polynomial applied, within reasonable limits. 

The index measurements made on the global composite spectrum are summarized in Table~\ref{tab:sigsplit}, and are 
compared to the model predictions in Figure~\ref{fig:biggx_indices}. 
The index errors are obtained by performing the index measurement on the bootstrap realisations of the composite spectrum.

The measured FeH0.99 is consistent with the Chabrier or Salpeter IMF for models with solar abundance mixtures. 
For higher assumed $\alpha$/Fe ratios, more dwarf-enriched models are brought into closer agreement with the data. 
For the mildly $\alpha$-enhanced model favoured by the optical data, the measured 
FeH0.99 is consistent with a Salpeter IMF, marginally inconsistent with Chabrier (at the 2.2$\sigma$ level) and incompatible with the $x$\,=\,3 model (at $>$3$\sigma$). 
Only by adopting [$\alpha$/Fe]\,=\,+0.4 (which is inconsistent with [Mg/Fe] measured from optical spectroscopy) can the $x$\,=\,3 model be brought 
within 2$\sigma$ of the observed index. 
Not unexpectedly, the predicted FeH0.99 is sensitive to the iron abundance, but this is measured to be close to solar from the 
fit to the optical indices. The index also depends on the sodium abundance, which is not constrained from our optical spectra. 
For the substantial sodium enhancements (up to +1.0\,dex) suggested by the results of CvD12b, the predicted FeH0.99 could be reduced by 0.2\,dex, 
which would yield agreement with the $x$\,=\,3 model.

As noted above, casual examination of Figure~\ref{fig:compo_biggx_imf} might suggest a better match for Chabrier models than for Salpeter, in apparent
disagreement with the index measurements. 
In fact, the index measurements implicitly force
a more local continuum matching than the low-order polynomial correction we applied for the spectral comparison. Because the models lie lower 
than the observed spectrum in the red pseudo-continuum band of FeH0.99, the predicted index is reduced 
relative to the observed value. Figure~\ref{fig:indexspec} (left) demonstrates this by plotting the WFB region after imposing a linear correction to normalise the 
pseudo-continua to unity in models and data alike. With this normalisation it becomes clearer how Salpeter (or slightly more dwarf-enriched) models yield
the closest agreement in the index value. However, this figure also highlights the failure of all models to match the detailed shape of the observed 
spectrum through the WFB region\footnote{The same discrepancy in the red pseudo-continuum of FeH0.99 is seen in the van Dokkum \& Conroy (2012) sample (their figure 8).}.

The measured CaI1.03 index for the observed composite spectrum is consistent with the Chabrier IMF, for either
the solar abundance ratio or the $\alpha$-enhanced models, and marginally inconsistent (2$\sigma$ with the Salpeter models. 
The dwarf-enriched $x$\,=\,3 models are excluded at the $>$5$\sigma$ level (see also Figure~\ref{fig:indexspec_cai}, left).
However, the CvD12a $\alpha$-enhanced models assume that calcium, being an $\alpha$-element, 
is enhanced in lock-step with magnesium and titanium etc. Optical spectroscopic indices (primarily Ca4227) imply instead that the Ca/Fe
ratio is lower than Mg/Fe on average (e.g. Vazdekis et al. 1997; Thomas, Maraston \& Bender 2003; Smith et al. 2009), in which case the predicted CaI1.03 will be 
too high. For the $\sigma$\,$>$\,100\,\kms\ galaxies in our global composite, the contribution-weighted average [Ca/Fe] is +0.07, compared to [Mg/Fe]=+0.25.
We estimate the effect of this by noting that for a Chabrier IMF (for which CvD12a provide a variety of models with variation in individual element abundances),
an increase by 0.15\,dex change in Ca/H increases the CaI1.03 index by 0.03\,\AA. 
Applying this as a linear correction to the indices predicted by the [$\alpha$/Fe]=+0.2 model, 
the Chabrier and Salpeter models are both at consistent with the data at the $\sim$1$\sigma$ level. 
The $x$\,=\,3 models would remain excluded ($>$4$\sigma$)\footnote{
Strictly of course, the low Ca/Mg ratios were derived for Salpeter IMF in the comparison with Schiavon (2007) models, rather than self-consistently with varying IMF. 
Figures 13 and 14 of CvD12a
tend to confirm that Ca4227 and Mgb5177 are much less sensitive to the IMF than to abundance variations, so this should not cause any serious bias.}.

We conclude that for Coma red-sequence galaxies with  $\sigma$\,$>$\,100\,\kms, the strengths of the Wing--Ford band and the IMF sensitive Ca\,{\sc i} line
at 10345\,\AA\ each support an IMF with dwarf-star content similar to the Salpeter case, once the known abundance information is taken into account. 
The concordance between the two indicators suggests that the average Na/Fe ratio, which influences the WFB, cannot be too large: 
[Na/Fe]$\,=$\,+1.0 would violate the consistency between FeH0.99 and CaI1.03, for {\it any} IMF, at the $>$2.5$\sigma$ level. 


\subsection{Dependence on velocity dispersion}\label{sec:sigmabin}

Most properties of red-sequence galaxies correlate with proxies for their mass, and in particular with internal velocity dispersion, $\sigma$. 
In general, galaxies with larger $\sigma$ have older SSP-equivalent ages and higher abundances of $\alpha$-elements (including magnesium, calcium, titanium), 
and of carbon and nitrogen, although the iron abundance is flat or even declining with $\sigma$ (e.g. Nelan et al. 2005; Graves et al. 2007; Smith et al. 2009; 
Johansson et al. 2012). 
Recent work (CvD12b; Spiniello et al. 2012; Ferreras et al. 2012) suggest that the IMF may also depend on galaxy mass. For example, CvD12b find
Chabrier-like IMFs favoured on average at $\sigma$\,$\approx$\,150\,\kms, Salpeter-like at $\sigma$\,$\approx$\,250\,\kms, and steeper still 
for the most massive galaxies. 
In this section we divide the FMOS sample according to velocity dispersion to examine the mass-dependence of the indices and derived IMF for our sample. 

In Figure~\ref{fig:compo_sigsplits}, we show composite infrared spectra computed for galaxies in four ranges of velocity dispersion. 
The bin boundaries were defined to yield a similar signal-to-noise ratio of $\sim$90\,\AA$^{-1}$ in each stack. 
For these smaller samples, the stacked spectra show occasional noise spikes which are probably
due to incomplete rejection of sky-subtraction residuals. It is clear that four is as many bins as can be justified by the quality of the present data.
The effective mean velocity dispersion, 
age and abundances (weighted according to the contribution of each galaxy to the stack) for each bin are given in Table~\ref{tab:sigsplit}.
Qualitatively, we find only weak variations in the spectra as a function of $\sigma$, with all the main features of the spectra reproduced in all four composites. 
There is a hint for slightly stronger WFB in the most massive bin, but lower in the second-most massive bin, and hence no clear trend is apparent. 
The low-$\sigma$ galaxies appear to show stronger Ca{\,\sc i}.

Figure~\ref{fig:indices_sigsplits} shows the FeH0.99 and CaI1.03 indices measured from the $\sigma$-binned stacks, with comparison to the global composite and
to the CvD12a model predictions.  Direct comparisons of observed and model spectra for the highest-$\sigma$ bin are shown in 
Figures~\ref{fig:indexspec} (right) and \ref{fig:indexspec_cai} (right) for FeH0.99 and CaI1.03 respectively.
The predicted indices are derived after broadening the models to the contribution-weighted velocity dispersion of each stack, i.e. the 
doppler broadening correction is applied to the models, rather than to the data.
For the predicted indices, we track the variation in $\alpha$/Fe ratio as a function of $\sigma$ by interpolating between the CvD12a models, 
with [$\alpha$/Fe]\,=\,0.0,\,+0.2,\,+0.3 and +0.4, to match the weighted-mean observed Mg/Fe. 
For FeH0.99, the models are adjusted to track the average measured Fe/H in each bin, and the predictions for CaI1.03 account for the average measured 
Ca/Fe\footnote{These differential corrections were derived from Chabrier IMF models, and applied to all other IMFs.}.

The behaviour of the measured FeH0.99 index in Figure~\ref{fig:indices_sigsplits} is not monotonic, with an apparently significant dip 
in the $\sigma$\,$\approx$\,190\,\kms\ bin, 
as expected from the spectra. Comparison to the constant-IMF model predictions shows that this behaviour is expected, 
once allowance is made for the variation in Fe/H (by chance, this bin has lower average iron abundance than any other). 
With this correction, all bins with velocity dispersion $<$210\,\kms\  are within 1$\sigma$ from the predictions consistent with a common Salpeter IMF. The 
highest velocity dispersion bin has slightly higher FeH0.99 (1.5$\sigma$ from the Salpeter prediction), 
but is still marginally inconsistent (2$\sigma$) with the $x$\,=\,3 model.

The CaI1.03 index is consistent with Salpeter models across the whole range in $\sigma$, 
once corrected for the measured Ca/Fe. (If instead we were to assume Ca/Fe\,=\,Mg/Fe, the results favour a Chabrier IMF.)
There is no evidence in this index for increased dwarf content at high velocity dispersion:
the $\sigma>210$\,\kms\ bin is consistent with Salpeter or Chabrier IMF, and inconsistent with the 
$x$\,=\,3 model at the 3.6$\sigma$ level.


As discussed above, the conclusions for the WFB are somewhat sensitive to the sodium abundance which is not well constrained from other data.
If we had assumed that Na/Fe increases with velocity dispersion, then we would have recovered a trend of increasing dwarf-enrichment with $\sigma$ from FeH0.99,
though CaI1.03 would be unaffected.

\subsection{Correlation with Mg/Fe ratio}\label{sec:mgfebin}

\begin{figure*}
\includegraphics[angle=270,width=178mm]{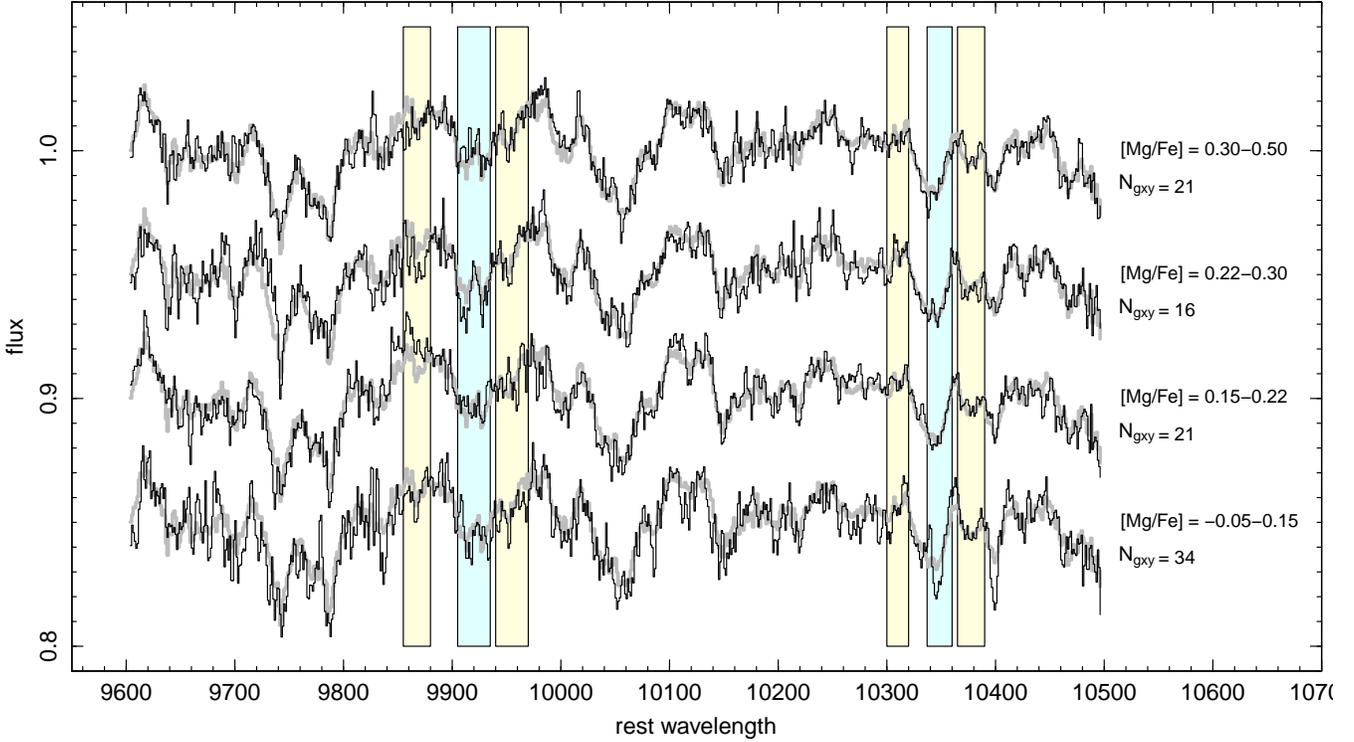}
\caption{Composite spectra computed in four bins of Mg/Fe (black lines), for galaxies with $\sigma$\,$>$\,50\,\kms. In each case, the thick grey line shows the 
fiducial composite of $\sigma$\,$>$\,100\,\kms\ galaxies for comparison. In this figure, no smoothing has been applied to either the Mg/Fe-split or the 
global composite.}
\label{fig:compo_mgfesplits}
\end{figure*}

\begin{figure*}
\includegraphics[angle=270,width=178mm]{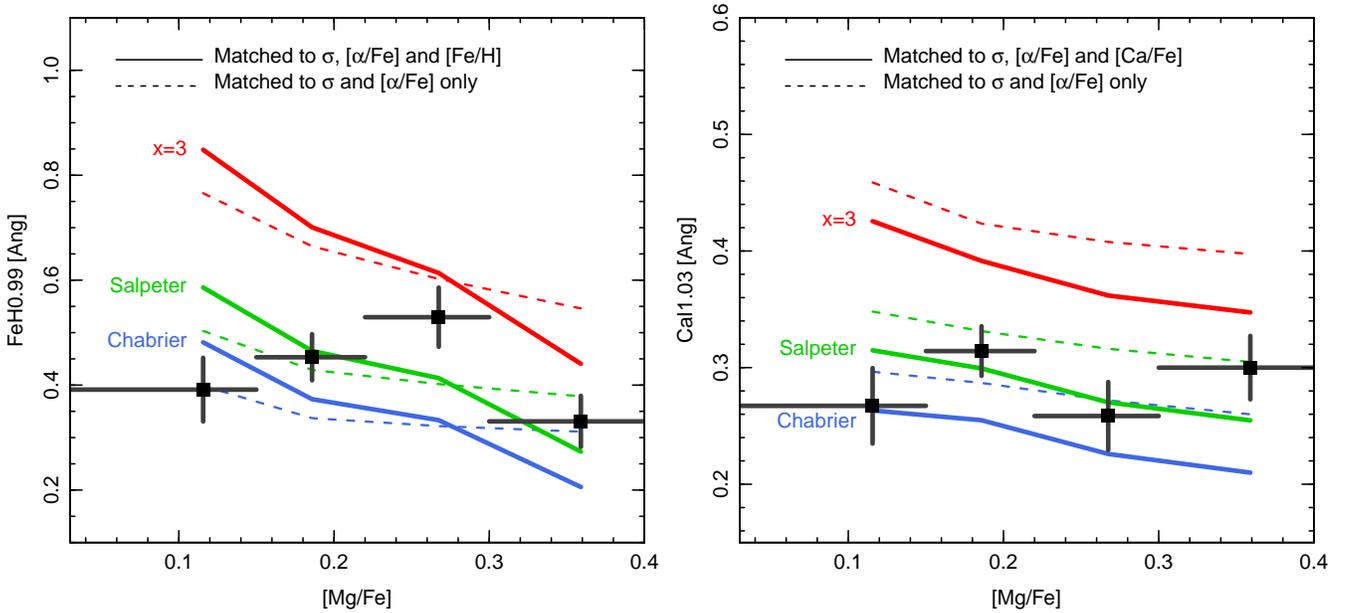}
\caption{Variation in the FeH0.99 and CaI1.03 indices as a function of [Mg/Fe]. Bins in [Mg/Fe] are as in Figure~\ref{fig:compo_mgfesplits}.
The models shown are as in Figure~\ref{fig:indices_sigsplits}.
}
\label{fig:indices_mgfesplits}
\end{figure*}

CvD12b find that the apparent increase in dwarf content in massive ellipticals may be better correlated with the Mg/Fe ratio, rather than with velocity dispersion
(their Figure 5). Such a correlation would be very interesting, as Mg/Fe is generally interpreted as an indicator for the timescale of star-formation (Thomas et al. 2005), 
and thus reflects the physical conditions in galaxies during their major formation epoch.

We test this result by binning our spectra according to (optically-determined) Mg/Fe (Figures~\ref{fig:compo_mgfesplits}-\ref{fig:indices_mgfesplits} 
and Table~\ref{tab:sigsplit}). For the WFB we measure an increase in the FeH0.99 index strength with increasing Mg/Fe, for the bins with [Mg/Fe]\,$<$\,0.3.
The models show that for constant IMF we would instead expect the index to fall with increased Mg/Fe (reduced Fe/H and increased velocity dispersion, 
as well as the direct effect of 
$\alpha$/Fe). The inferred IMF drifts from Chabrier (or lighter) at [Mg/Fe]\,=\,+0.12 to Salpeter at [Mg/Fe]\,=\,+0.19 and close to $x$\,=\,3 at [Mg/Fe]\,=\,+0.27.
For the [Mg/Fe]$>$0.3 bin, a lower  index value is measured, but the recovered IMF would still be marginally heavier than Salpeter for the Fe/H-corrected models.
The CaI1.03 index shows similar behaviour: the index value is fairly constant with Mg/Fe, within the errors, but fixed-IMF models predict 
a declining trend (mainly due to increased velocity broadening as the high-Mg/Fe bins have higher $\sigma$ on average). Hence the inferred IMF
becomes more dwarf-enriched with increasing Mg/Fe, ranging from Chabrier to slightly heavier than Salpeter. 

The observed trends with Mg/Fe are clearly only marginally significant in either index individually. 
Taken in combination, however, the similar pattern derived from the two indices lends tentative support to the claim that 
$\alpha$-enhanced populations formed in rapid bursts harbour more low-mass stars than galaxies with extended star-formation histories. 

%
%

\section{Discussion}\label{sec:discuss}

In this section we focus on comparing our results to those presented (during completion of this paper) by Conroy \& van Dokkum (CvD12b). 
Their study is directly comparable to ours in probing a similar range in galaxy properties (although they have only two or three galaxies in the 
range covered by our lowest $\sigma$ bin of 50--130\,\kms), and in using optical spectra to constrain additional parameters (age, element abundances) 
so that possible IMF variations can be investigated. 

Our analysis differs from that of CvD12b first in combining the data from many galaxies into composite spectra. This has the advantage of rejecting (or at least
smoothing over) sky-subtraction residuals and other artefacts, but obviously limits the degree to which we can 
test IMF variations within the sample. Second, our analysis method is different from that of CvD12b: whereas they perform a full-spectrum fit (with a total of 
22 free parameters), over their optical and far-red spectra simultaneously, we instead adopt most parameters from the best-fitting SSP model in the optical, 
and focus on the infrared indices to isolate the possible IMF variations. 

A novel aspect of our work is that we introduce and 
exploit a new IMF-sensitive index measuring the Ca{\,\sc i} line at 10345\,\AA. In contrast to {\it all} of the ``classical'' IMF-sensitive features (WFB, Na{\,\sc i} doublet, 
Ca{\,\sc ii} triplet), this index is nearly independent of the sodium abundance, and hence provides an independent test which complements results obtained 
from other features.
We quantify this statement using the CvD12a models: Using the 13.5\,Gyr, solar-abundance, Chabrier-IMF case as a fiducial point, 
we find that increasing Na/Fe by 0.3\,dex increases the NaI0.82 index by 0.14\,\AA. If instead the Na/Fe is held fixed, but the IMF is changed from Chabrier 
to Salpeter, this index increases by 0.12\,\AA. The ratio of sodium sensitivity to IMF sensitivity  for NaI0.82 is thus 0.14\,/\,0.12\,=\,+1.17.
For the CaII0.86 index (measuring the Ca{\,\sc ii} triplet), the equivalent ratio is +0.44, while for FeH0.99 (measuring the WFB), the ratio is --0.56. 
(As emphasized by CvD12a, sodium indirectly affects the strength of many spectral features, since it is one of the dominant electron donors
in cool stellar atmospheres, and hence changes the ionisation balance of other species.)
By contrast, increasing Na/Fe by 0.3\,dex reduces the CaI1.03 index by 0.002\,\AA, while changing to a Salpeter IMF increases this index by 
0.036\,\AA\ and hence the sensitivity ratio is --0.06, an order of magnitude smaller than the classical indices.

Our first-order result is that the WFB and Ca{\,\sc i} measurements favour a Salpeter-like IMF on average in red-sequence/early-type galaxies, 
rather than the bottom-light Chabrier/Kroupa form found in the Milky Way. This is an important result, since it implies  a $\sim$50\,per cent correction to 
the K-band mass-to-light ratios for early-types relative to spirals. This conclusion is consistent with the average result from CvD12b in the relevant
mass range, and also agrees with recent lensing and dynamical analyses (Spiniello et al. 2011; Cappellari et al. 2012). 

We do not recover a clear signal of {\it increasing} dwarf-enrichment with increasing mass (traced by velocity dispersion). The evidence for 
such a trend in CvD12b is driven mainly by (a) their original stacked spectra of four Virgo galaxies with $\sigma$\,$\approx$\,300\,\kms\ from van Dokkum \& Conroy (2010),
which require an IMF significantly heavier Salpeter, and (b) their $\sim$10 galaxies at  $\sigma$\,$\la$\,180\,\kms\ which favour a Chabrier-like IMF on average. 
We have few galaxies in the 300\,\kms\ regime, and our highest-$\sigma$ bin is dominated by galaxies with $\sigma$\,$\approx$\,240\,\kms. 
Hence, our results do not exclude a steeper IMF at the high-mass extreme. 
At lower $\sigma$, however, where our results are based on a much larger galaxy sample, our data clearly favour a Salpeter IMF, in apparent conflict with the 
Chabrier-like solutions recovered by CvD12b. 

Our analysis supports the trend found by CvD12b for an increase in the dwarf content of red-sequence galaxies as a function of their 
$\alpha$-abundance ratio. 
Notably, the only galaxies for which we infer a Chabrier-like IMF, similar to that of the Milky Way, are those which have abundance patterns 
closest to that of the Milky Way ([Mg/Fe]\,$\approx$\,0.1).
For galaxies with high $\alpha$-enhancements ratios ([Mg/Fe]\,$\ga$\,0.3), thought to arise from formation 
in very rapid and intense star-bursts, the indices require IMFs that are heavier than Salpeter, though not so extreme as an $x$\,=\,3 
power-law\footnote{Our results refer strictly to the IMF at low stellar mass ($\la$\,0.5\,$M_\odot$). Although a top-heavy IMF 
in bursts has been proposed 
as an explanation for the sub-millimeter galaxy counts (Baugh et al. 2005), this needs only to affect the IMF at much higher masses, 5--20\,$M_\odot$
(see Lacey et al. 2010). Hence it is possible to construct an IMF satisfying both constraints, 
although we reserve judgement as to whether it is physically plausible.}.

We emphasize that results obtained from the two measured indices are consistent with each other. CvD12b note that their results
depend somewhat on which combinations of IMF-sensitive features they include in their fits (although their overall conclusions are robust). 
For example if the Na\,{\sc i} region is removed from the fit, they recover lower dwarf-enrichments, whereas if both Na{\,\sc i} and the Ca{\,\sc ii} triplet are removed, 
so that only the WFB is constraining the IMF, then even heavier solutions are recovered. Their best-fitting models require large enhancements 
in Na/Fe, of up to an order of magnitude higher than the solar ratio in some cases. Since all three classical IMF indicators are affected 
(directly or indirectly) by the sodium abundance, 
it is important for future work to determine whether this high sodium enrichment is real, and not an artefact of inconsistencies between the various fitted
features\footnote{Enhancements of 0.5--1.0\,dex in sodium have been reported for high-metallicity Milky Way bulge giants (Cunha \& Smith 2006; Lecureur et al. 2007), 
but only mild ($\sim$0.1\,dex) enhancements were found by Bensby et al. (2010) for bulge dwarfs and subgiants.}. In the meantime, it is encouraging that our
sodium-independent CaI1.03 index yields results which are concordant with those from the WFB.

We finish by briefly comparing our results to two recent spectroscopic works based on the NaI doublet. Spiniello et al. (2012) analysed galaxies with 
velocity dispersions generally larger than average for our sample, but overlapping our two highest-$\sigma$ bins. 
They find a trend of increasing IMF slope from Salpeter-like at $\sigma$\,$\approx$\,200\,\kms\ (in agreement with our results at the same velocity dispersion)
to $x$\,$\approx$\,3 at $\sigma$\,$\approx$\,335\,\kms\ (a regime not probed by our sample). Ferreras et al. (2012) report consistency
with a Salpeter-like IMF at  $\sigma$\,$\approx$\,200\,\kms and  $x$\,$\approx$\,3 at 300\,\kms. At lower mass, they infer a constant Kroupa-like IMF at 
$\sigma$\,$\la$\,150\,\kms, which may be marginally discrepant with the Salpeter IMF favoured by our results in this mass range. 
This comparison should be viewed with caution however, since Ferreras et al. have not yet taken into account the possible
effect of Na/Fe variations as a function of $\sigma$. We note that our sample differs from both Spiniello et al. and Ferreras et al. in studying
only rich galaxy cluster members. It is not yet known whether apparent IMF variations are correlated in any way with galaxy environment.

\section{Conclusions}\label{sec:concs}

We have presented new infrared spectroscopy for red-sequence galaxies in Coma, 
and measured gravity-sensitive absorption features to probe their low-mass stellar content. 
Compared to other recent work in this field, our study differs in targetting cluster galaxies, and in sampling ``typical'' red-sequence
galaxies, rather than only the most massive objects.
By comparing line-strength indices against state-of-the-art spectral synthesis models, tuned to match 
element abundances estimated from optical data, we derive constraints on the average IMF and its variation with velocity dispersion
and $\alpha$-abundance ratio. 

Our main conclusions are as follows:
\begin{itemize} 
\item The average rest-frame 1$\mu$m spectra for sizable samples of galaxies at the distance of Coma
can be recovered with Subaru/FMOS, thanks to its wide-field multiplex capability which is currently unique in the infrared. 
\item The observed spectral features are generally in good qualitative agreement with the latest stellar libraries and synthesis models, although
some localised discrepancies are seen. 
\item The Ca{\,\sc i} line at 10345\,\AA\ is promising as a new gravity-sensitive feature, which unlike the ``classical'' IMF indicators is independent of the sodium abundance.
\item The strength of the Wing--Ford band and Ca{\,\sc i} 10345\,\AA\  line indicate that red-sequence galaxies in Coma  have (on average) 
a dwarf-star content similar to that in a Salpeter IMF. 
\item There is no clear trend in the derived IMF as a function of velocity dispersion, with Salpeter models an adequate fit from $\sigma$\,$\approx$\,100\,\kms\ to $\sigma$\,$\approx$\,250\,\kms.
\item A more dwarf-dominated IMF cannot be ruled out at the highest velocity dispersions ($\sigma$\,$\ga$\,300\,\kms), which are not well sampled by our observations.
\item The derived IMF is correlated with Mg/Fe ratio suggesting that galaxies which underwent intense rapid starbursts formed a larger number of
low-mass stars per solar-mass star. 
\end{itemize}

Our work adds to a  developing consensus that massive red-sequence galaxies have heavier IMFs, with a greater contribution from low-mass stars, 
than found in the Milky Way. A Salpeter-like IMF is favoured by recent spectroscopic (CvD12b; Spiniello et al. 2012; Ferreras et al. 2012; this work), 
dynamical (Thomas et al. 2011; Cappellari et al. 2012) and lensing analyses (Spiniello et al. 2011; Sonnenfeld et al. 2012). 
For spiral galaxies, in contrast, recent lensing results are consistent with Chabrier-like IMFs similar to the Milky Way (Brewer et al. 2012), consistent with 
earlier dynamical results (Bell \& de Jong 2001). Together with the apparent correlation of dwarf-star content with Mg/Fe ratio found here and in 
CvD12c, these results strongly suggest that the low-mass IMF may be dictated in part by the dominant mode of star-formation: 
quiescent star formation yields Kroupa-like IMF, while star-bursts lead to an excess of low-mass stars.

\section*{Acknowlegments}

We are grateful to Kentaro Aoki for expert assistance in our Subaru observations and to Charlie Conroy for providing model spectra in advance of publication.
RJS was supported for this work by STFC Rolling Grant PP/C501568/1 ``Extragalactic Astronomy and Cosmology at Durham 2008--2013''.

{}

\label{lastpage}

\end{document}